\documentclass[floats]{revtex4}

\usepackage{graphicx}

\begin{document}

\title {Spectral Function in Mott Insulating Surfaces}

\author{L. O. Manuel$^1$, C. J. Gazza$^1$, A. E. Feiguin$^2$ and
A. E. Trumper$^1$}

\affiliation {$^1$Instituto de F\'{\i}sica Rosario (CONICET) and
Universidad Nacional de Rosario,
Boulevard 27 de Febrero 210 bis, (2000) Rosario, Argentina\\
$^2$Department of Physics and Astronomy, University
of California, Irvine, California 92697, USA}

\vspace{4in }

\date{\today}

\begin{abstract}
We show theoretically the fingerprints of short-range spiral magnetic correlations
in the photoemission spectra of the Mott insulating ground states realized in the
$\sqrt{3}\times\sqrt{3}$ triangular silicon surfaces K/Si(111)-B and SiC(0001).
The calculated spectra present low energy features of magnetic origin with a reduced
dispersion $\sim 10-40$ meV compared with the center-of-mass spectra
bandwidth $\sim 0.2-0.3$ eV.
Remarkably, we find that the quasiparticle signal survives only around the magnetic
Goldstone modes.
Our findings would position these silicon surfaces as new
candidates to investigate non-conventional quasiparticle excitations.
\end{abstract}

\maketitle

It is well known that the strong correlation in electronic systems
with an odd number of electrons per unit cell may give rise to
Mott insulating (MI) ground states, in contradiction with band
theory predictions. Generally, the presence of metal-transition
ions with unpaired electrons are responsible for the MI properties
of many compounds like the undoped high-$T_c$ cuprates. However,
it has also been shown that surfaces characterized by a
$\sqrt{3}\times\sqrt{3}$ arrangement of silicon dangling-bond
surface orbitals can provide the ideal conditions to realize the
MI phenomena without metal-transition
ions\cite{weitering,johansson96,tosatti,northrup}. Although the
on-site Coulomb potential $U$ is not so large, the surface state
bandwidth is reduced due to reconstructions, becoming the
correlation effect important. Recently, Weitering {\it et
al.}\cite{weitering} used moment-resolved direct and inverse
photoemission spectroscopy to demonstrate that the triangular
interface K/Si(111)-$(\sqrt{3}\times\sqrt{3})$-B (hereafter
K/Si-B) has a MI ground state. Even though such an insulating
state is not necessarily unstable toward antiferromagnetism due to
the lack of perfect nesting, it has also been argued that K/Si-B
is the first experimental realization of a triangular Heisenberg
spin-1/2 model with a 120$^{\circ}$ N\'eel order. However,
standard density-functional methods combined with exact
diagonalization studies suggest a Mott insulator with a
non-magnetic ground state\cite{hellberg}. In addition, it was
speculated that the reconstructed triangular surface
SiC(0001)-$(\sqrt{3}\times\sqrt{3})$ \cite{johansson96} (hereafter
SiC) also presents a $120^{\circ}$ N\'eel ordered ground
state\cite{tosatti}. Since direct measurements of magnetic order
in such surfaces are difficult to implement, the magnetic
properties of these MI ground states still remain
unanswered\cite{tosatti}. On the other hand, nowadays it is
possible to perform higher resolution photoemission experiments
than the previous ones, which would allow a detailed analysis of
the single-hole properties in these MI ground states. These issues
motivated us to study theoretically the single-hole dynamics in a
triangular antiferromagnet (AF) with two objectives: i) to
investigate the effect of a frustrated magnetic order in the
quasiparticle (QP) behavior, and ii) to obtain spectroscopic
fingerprints of magnetic order through the photoemission spectra
calculated for realistic parameters of the surfaces K/Si-B and
SiC. A careful comparison with future higher resolution
spectroscopy experiments could give some information about the
underlying short-range magnetic structure.

As a consequence of the magnetic order,
we find a strong k-dependence of the spectral function structures.
In particular, we show the emergence of spectral weight of magnetic
origin between the Fermi level and the measured surface state band, with
a strongly reduced dispersion.
For the surfaces  K/Si-B and SiC, our results
show a remarkable vanishing of the QP weight for a large region of the
Brillouin zone (BZ) outside
a neighborhood of the magnetic Goldstone modes.

To tackle this  problem we assume the $t\!-\!J$ model, which is supposed
to capture the low-energy physics involved in photoemission spectroscopy
of antiferromagnetic Mott insulators\cite{wells}.
In order to solve the model,
we use the self-consistent Born approximation (SCBA) complemented with
exact diagonalization studies.
For non-frustrated AF, it has already been
established the success of the SCBA to reproduce exact diagonalization
results on small clusters\cite{martinez} and angle-resolved
photoemission spectroscopy (ARPES) experiments\cite{wells}.

We assume a surface ground state
with a magnetic wave vector ${\bf Q}=(4\pi/3,0)$ lying in the surface
plane $x\!-\!z$, and spin waves as the low energy excitations.
We perform a unitary transformation to local quantization axis so
as to have a ferromagnetic ground state in the  $z^{\prime}$
direction.
Then, we use the spinless fermion
$\hat{c}^{\prime}_{i \uparrow}=h^{\dagger}_i$, $\;\;
\hat{c}^{\dagger \prime}_{i \downarrow}=h_i S^{-}_i$,
and the Holstein-Primakov
$S^{x^\prime}_i \sim \frac{1}{2}(a^{\dagger}_i+a_i)$,
$S^{y^\prime}_i \sim \frac{{\it i}}{2}(a^{\dagger}_i-a_i)$,
$S^{z^\prime}_{i} = \frac{1}{2}-a^{\dagger}_i a_i$ representations.
These are replaced in the $t\!-\!J$ model keeping the relevant terms
up to third order. After a lengthy but straightforward calculation, the
Hamiltonian results

\begin{eqnarray}
\label{HSW}
H  =  \sum_{\bf k} &  \epsilon_{\bf k}h^{\dagger}_{\bf{k} } h_{\bf{k} } &
 + \sum_{\bf q}  \omega_{\bf q} \alpha^{\dagger}_{\bf q} \alpha_{\bf q}
\nonumber  \\
&  - t\sqrt{\frac{3}{N_s}} & \sum_{\bf k, q}
\left[M_{\bf kq}h^{\dagger}_{\bf k}
h_{\bf k-q}\alpha_{\bf q} + h.c. \right].
\end{eqnarray}
In the Hamiltonian (\ref{HSW}), $\epsilon_{\bf k}=-t\gamma_{\bf
k}$ and $\omega_{\bf q}=\frac{3}{2}J\sqrt{(1-3\gamma_{\bf
q})(1+6\gamma_{\bf q})}$ are the hole and magnon dispersions,
respectively. $\gamma_{\bf k}=\sum_{{\bf {\delta}}}\cos({\bf
k}.{\bf {\delta}})$ and $\beta_{\bf k}=\sum_{ \bf{\delta}}
\sin({\bf k}.{\bf{\delta}})$ are geometric factors, with ${\bf
\delta}$ the positive vectors to nearest neighbors, and ${\bf k}$
varying in the first BZ of the $\sqrt{3}\times\sqrt{3}$ surface.
The bare hole-magnon vertex interaction is defined by $M_{\bf
kq}=i (\beta_{\bf k}v_{-\bf{q}} -\beta_{\bf k-q}u_{\bf q})$ with
the Bogoliubov coefficients $u_{\bf q}= \left[(1+3\gamma_{\bf
q}/2+ \omega_{\bf q})/2\omega_{\bf q}\right]^{\frac{1}{2}}$ and
$v_{\bf q}=$ sign$(\gamma_{\bf q})\left[(1+ \gamma_{\bf
q}/2-3\omega_{\bf q})/2\omega_{\bf q}\right]^{\frac{1}{2}}$. The
free hopping hole term implies a finite probability of the hole to
move without emission or absorption of magnons because of the
underlying {\it non-collinear} magnetic structure. The hole-magnon
vertex interaction adds another mechanism for the charge carrier
motion which is magnon-assisted. The latter is the responsible for
the spin-polaron formation when a hole is injected in a
non-frustrated antiferromagnet\cite{kane,martinez}. We will show
the existence of a subtle interference between both processes that
turns out to be dependent on the momenta. An important quantity
that allows us to study the interplay between such processes is
the retarded hole Green function that is defined as $G_{\bf
k}^{h}(\omega)= \langle AF| h_{\bf k}\frac{1}{(\omega+{\it
i}\eta^{+}- H )} h^{\dagger}_{\bf k} |AF\rangle$, where
$|AF\rangle$ is the undoped antiferromagnetic ground state with a
$120^{\circ}$ N\'eel order. In the SCBA the self energy at zero
temperature results
$$
\Sigma_{\bf k}(\omega )=\frac{3t^2}{N_s}\sum_{\bf q}
\frac{\mid M_{\bf kq}\mid^{2}}{\omega -\omega_{\bf q}-\epsilon_{{\bf k}-{\bf q}}
-\Sigma_{\bf k-q}(\omega-\omega_{\bf q})}.
$$
We have solved numerically this self-consistent equation for
$\Sigma_{\bf k}(\omega)$, and calculated the hole spectra function
$A_{\bf k}(\omega ) = - \frac{1}{\pi}$Im$G_{\bf k}^{h}(\omega)$ and
the quasiparticle (QP) weight $z_{\bf k}=\left(1-
\frac{\partial \Sigma_{\bf k}(\omega)}{\partial \omega}\right)^{-1}\vert_{E_{\bf k}}$
, where the QP energy is given by $E_{\bf k}=\Sigma_{\bf k}(E_{\bf k})$.
Before we discuss the results it is important to mention  that, unlike
previous works\cite{azzouz}, we will concentrate on the behavior
of the photoemission spectra for realistic parameters of  the surfaces
K/Si-B and SiC, what implies a strong coupling regime and a careful
extrapolation to the thermodynamic limit (we have studied cluster sizes up to
2700 sites).

 For the K/Si-B (SiC) surface, photoemission spectra
give a bandwidth $W \sim 0.3$ eV ($\sim 0.2$ eV) for the occupied
surface electronic band\cite{weitering,johansson96}.
These experiments together with inverse
photoemission indicate an effective on-site Coulomb repulsion
$U \sim 1\!-\!2$ eV ($\sim 1.5\!-\!2$ eV).
Electronic band calculations give a similar value for $U$ and a wider
surface bandwidth $W \sim 0.61$ eV ($\sim 0.35$ eV), due to the neglection
of correlation effects, leading to a {\it positive} $t \sim 0.07$ eV
($\sim 0.04$ eV) \cite{hellberg,northrup}.
The large $U/W$ ratios indicate the
presence of strong electronic correlations and a considerable
suppression of charge fluctuations. In addition, scanning tunneling
microscopy measurements on the SiC surface show a clear evidence
of a Mott insulating state\cite{ramachandran99}.
Theoretical works on the Hubbard model suggest that these surfaces
would be located
in the antiferromagnetic region of the phase diagram\cite{capriotti}.
All these estimations point out to $J/t \sim 0.1 -0.4$ for both surfaces.
In that parameter range, we have observed negligible quantitative
changes in our results, so we take $J/t=0.4$ as a reference value.

\begin{centering}
\begin{figure}[ht]
\vspace*{0.cm}
\includegraphics*[height=0.28\textwidth]{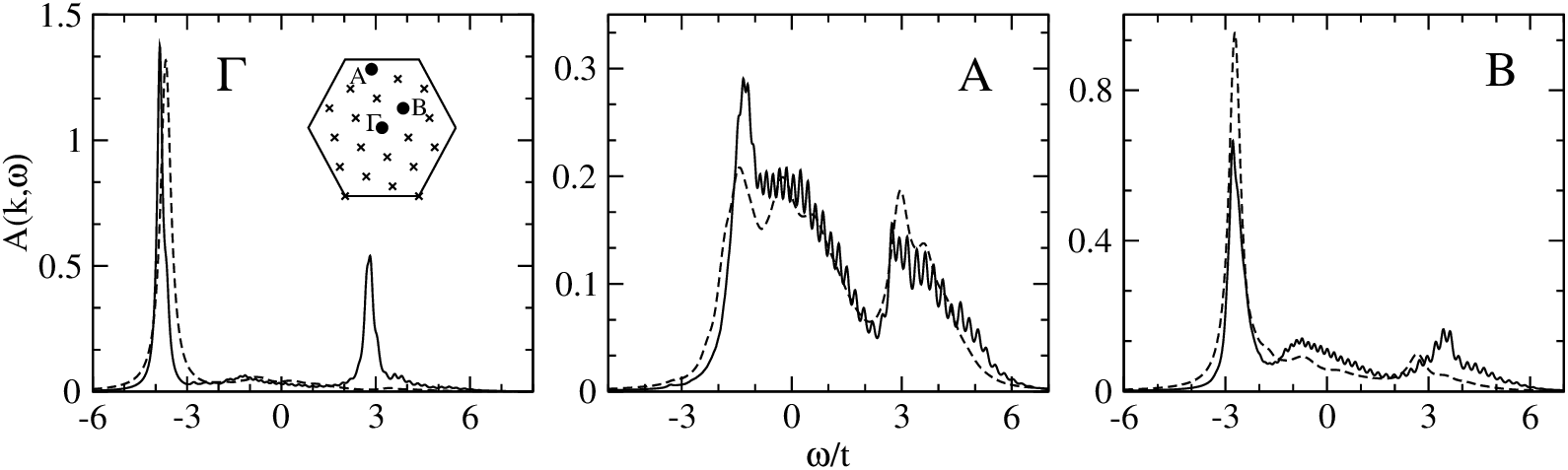}
\caption{Spectral functions versus frequency for  $J/t=0.4$ and
$N=21$ corresponding to the momenta $\Gamma=(0,0)$,
$A=\frac{4\pi}{21}(-1,3\sqrt{3})$ and $B=\frac{4\pi}{21}(2,\sqrt{3})$ shown
as filled circles in the inset of the left panel (the crosses
represent the other momenta).
The solid and dashed lines are the
exact and SCBA results, respectively.}
\label{spectral}
\end{figure}
\end{centering}

In order to test the validity of the SCBA we have compared its results
for the single-hole dynamics with Lanczos exact diagonalization (ED) calculations
on small clusters. In general, we have found a good agreement for all momenta
of the BZ and for $J/t \lesssim 1$.
In Fig. \ref{spectral} we show the spectral functions for a cluster size of $N=21$ and
in the strong coupling regime $J/t =0.4$, for three momenta of the BZ.
There is a very good agreement between ED and
SCBA results in the whole range of energies.
This concordance gives a strong support to our analytical approach and so,
from now on, we will focus on the SCBA spectra in the thermodynamic limit.

\begin{figure}
\includegraphics*[width=0.50\textwidth]{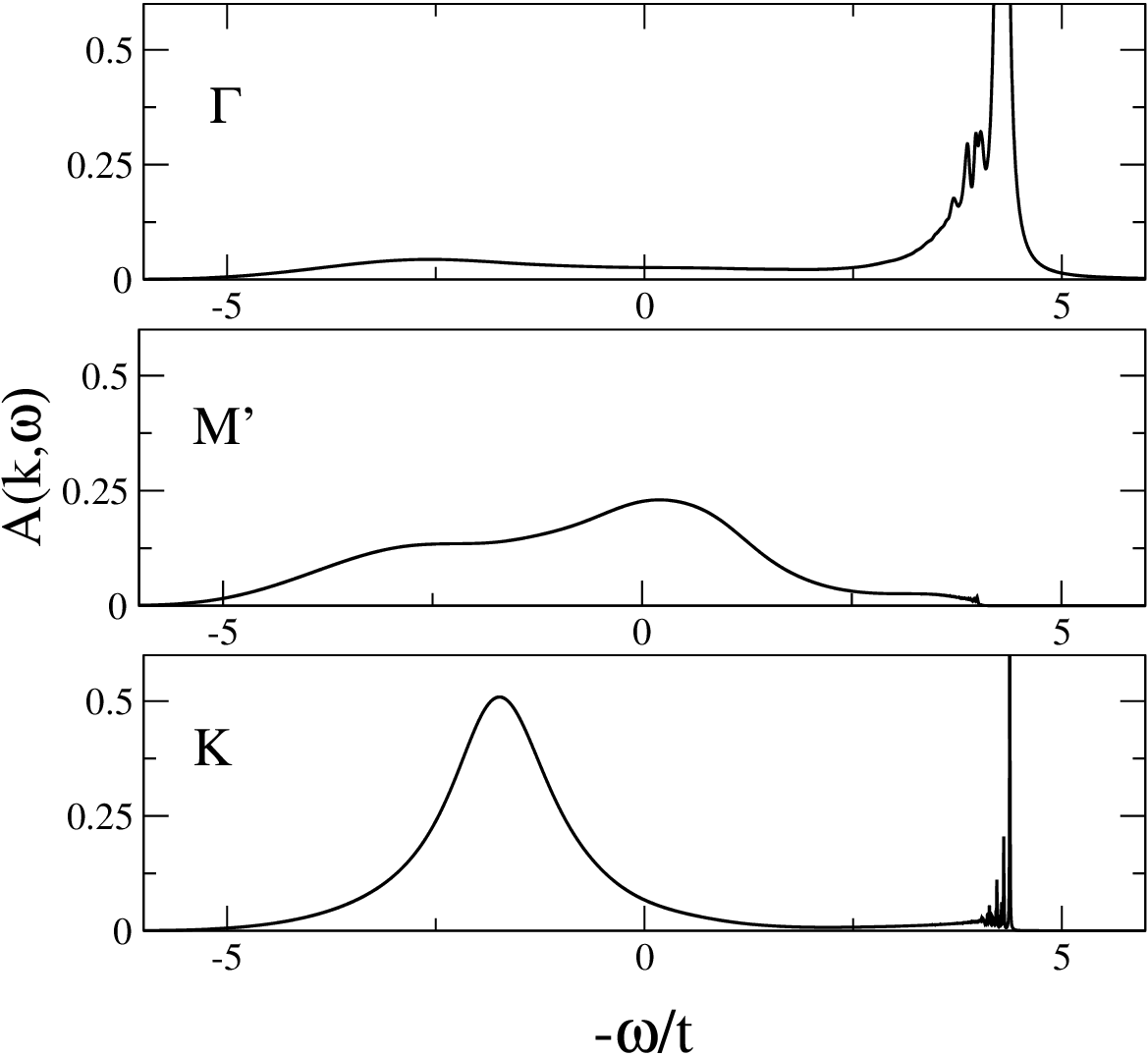}
\caption{
Spectral functions for realistic parameters of the silicon
surfaces ($J/t=0.4$) calculated at three different points of
the BZ (see inset Fig. \ref{mbz}). Notice that the
Fermi level has been placed on the right.}
\label{QP}
\end{figure}

In Fig. \ref{QP} we show the spectral function structures,
which can be traced back to the distinct hole motion processes mentioned above.
The spectra extend over a frequency range of $\sim 9t$, that is, the
non-interacting electronic bandwidth. Similar values has been found in the
experiments on the silicon surfaces\cite{weitering,johansson96}.
For momentum $\Gamma$ (upper panel) both processes contribute coherently
to the quasiparticle excitation at the top of the spectra.
There is also a small incoherent component centered around $6t$ below
the QP peak.
When moving away from $\Gamma$, there is a spectral weight transfer from
the low energy coherent sector to higher energies.
At $M^{\prime}$ (middle panel) the coherence is completely
lost, surviving only a magnetic tail at low energy and a structureless
background.
At $K$(lower panel), corresponding to the magnetic vector $\bf Q$,
the coherence emerges once again, giving rise to the QP.
In the background there is a broad resonance
related to the free hopping hole motion with a finite lifetime of order
$\sim 4J$.

\begin{figure}
\includegraphics*[width=0.40\textwidth]{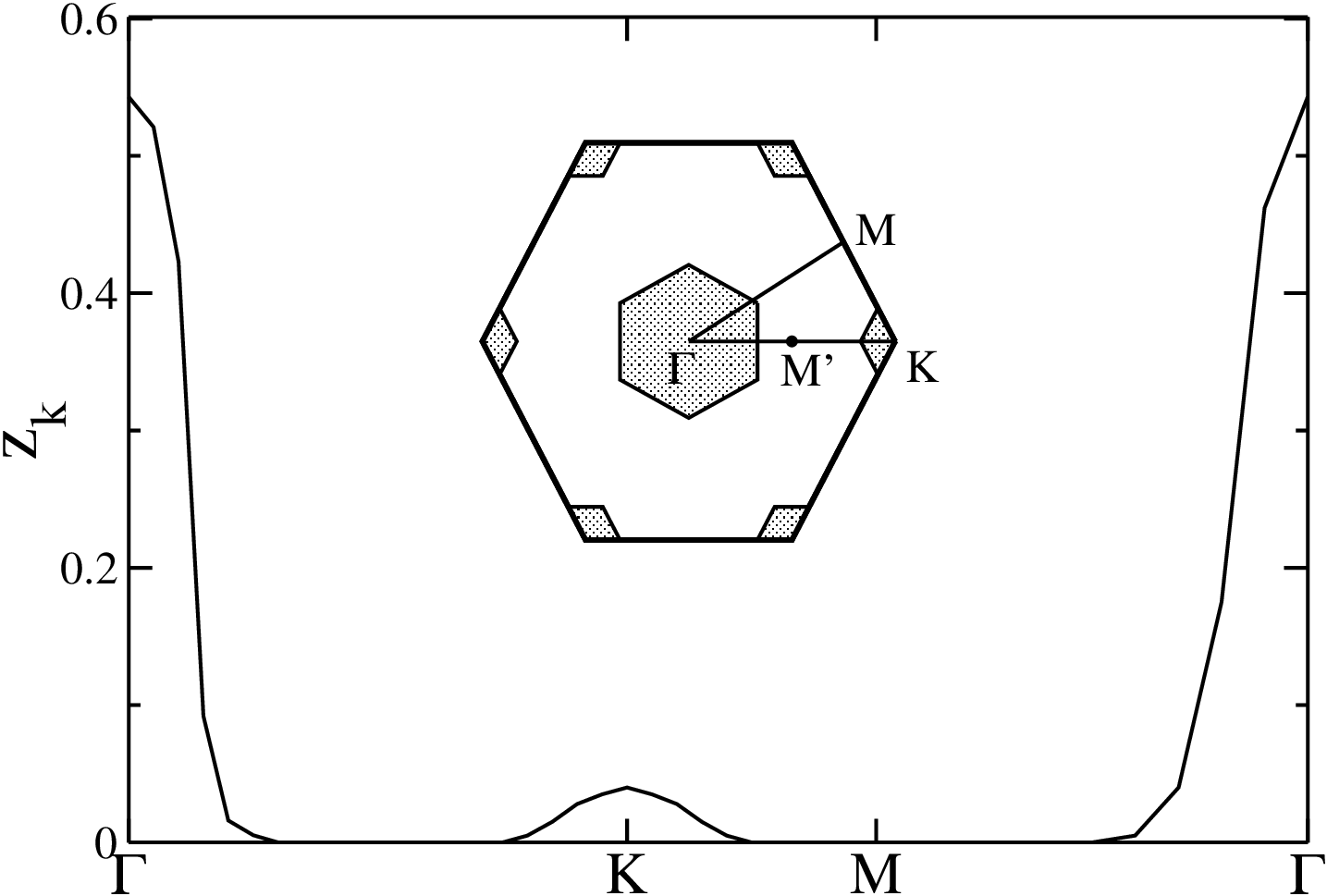}
\caption{ QP intensity along the $\Gamma-K-M-\Gamma$ path for
$J/t=0.4$. Inset:
 $\sqrt{3}\times\sqrt{3}$ Brillouin zone. In the shaded areas the QP weight is finite.}
\label{mbz}
\end{figure}

In Fig. \ref{mbz} we show the intensity of the QP peak along high
symmetry axes and, in the inset, the region of the BZ where the
quasiparticle weight is finite for $J/t=0.4$. Besides the strong
k-dependence, it can be observed that there is no QP for momenta
outside the neighborhood of the magnetic Goldstone modes (${\bf
k}={\bf 0}, {\bf Q}$). The existence of the quasiparticle peak is
also observed for an ample range of $J/t$ values. In particular,
as $J/t$ increases the shaded areas in the inset of Fig.
\ref{mbz}, where the quasiparticle weight is finite, increase and
only for values $J/t \gtrsim 2.5$ there is quasiparticle peak all
over the Brillouin zone. The non existence of quasiparticle is a
striking manifestation of the strong interference between the free
and magnon-assisted hopping processes. It is expected that this
feature could be observed in any system with short-range
non-collinear correlations.

\begin{figure}
\includegraphics*[width=0.40\textwidth]{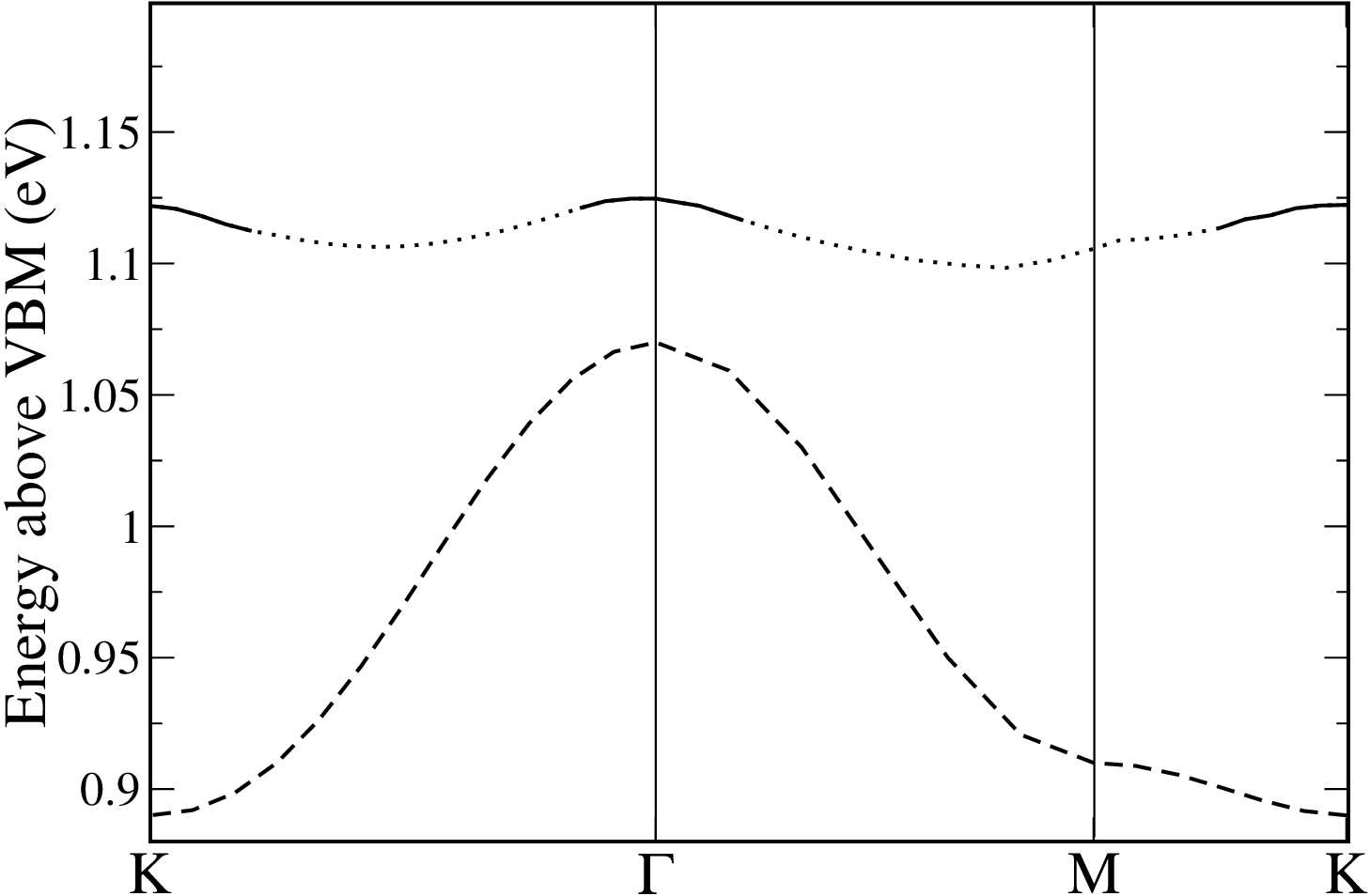}
\caption{ Surface band structure for realistic parameters
($J/t$=0.1, $t=0.04$ eV) of the
SiC(0001)-$(\sqrt{3}\times\sqrt{3})$ surface. The dotted line is
the photothreshold energy, the solid one is the QP dispersion. and
the dashed one is the center-of-mass spectra band. Energies are
given relative to the measured bulk valence-band maximum (VBM).
Experimentally the Fermi level is located at $2.3$ eV above the
VBM} \label{rel}
\end{figure}

The available photoemission spectroscopy data for the silicon surfaces
have a low energy resolution ($\sim 100\!-\!200$ meV) \cite{weitering,johansson96}
and it is impossible to discern the energy structure of the surface spectra.
The surface bands obtained experimentally are correctly reproduced
by the center-of-mass of our spectra.
The latter coincides with the
free hopping dispersion (see Fig. \ref{rel}) as it is expected from the exact
treatment of the first spectral moment of our theory.
Using the transfer integral $t$ obtained from
first-principles local-density-approximation calculations, \cite{northrup}
our bandwidths compare very well with the
experimental ones, reflecting the narrowing induced by the electronic
correlation. For the K/Si-B (SiC) surface we have obtained $W \sim 0.3$ eV
($\sim 0.18$ eV).
In  Fig. \ref{rel} we show the center-of-mass
and the photothreshold dispersions for the SiC surface.
When the quasiparticle exists, the photothreshold corresponds
to the QP energy excitation.
As it can be observed, the underlying magnetic structure changes the center-of-mass
minimum at the $K$ point to a QP energy
local maximum, nearly degenerate with the hole ground
state momentum $\Gamma$. There is also an appreciable reduced bandwidth $\sim 10$ meV
of the photothreshold dispersion in comparison with the measured surface
state bandwidth $\sim 200$ meV. Whereas for the K/Si-B surface the values are
$\sim 40$ meV and $\sim 300$ meV, respectively.

In conclusion, we have studied theoretically the hole spectral
function in the triangular $t\!-\!J$ model for realistic
parameters relevant for the silicon surfaces
SiC(0001)-$(\sqrt{3}\times\sqrt{3})$ and
K/Si(111)-$\sqrt{3}\times\sqrt{3}$. Assuming the presence of a
long-range magnetic N\'eel order we have observed the emergence of
low energy features of magnetic origin with a reduced dispersion
band. As the photoemission spectrum is not sensitive to the
asymptotic low energy magnetic properties of the system, we
speculate however that it could give important information about
the presence of short-range magnetic order. We have also obtained
an unexpected vanishing of the QP weight for a large region of the
Brillouin zone for these Mott insulating surfaces. Our theoretical
predictions could provide useful ground to the analysis of future
improved photoemission experiments. Using a simple and reliable
analytical method (SCBA), we have found clear signatures of
interesting physics caused by strong electronic correlation on
simple silicon surfaces.

This work was done
under PICT Grant No. N03-03833 (ANPCyT) and was partially supported by
Fundaci\'on Antorchas.


\begin{thebibliography}{99}
\bibitem{weitering} Weitering H H, Shi X, Johnson P D, Chen J, DiNardo N J
and Kempa K 1997 {\it \prl}{\bf 78} 1331
\bibitem{johansson96} Johansson L I, Owman F and M\aa rtensson P
                      1996 {\it Surf. Sci.} {\bf 360} L478
\bibitem{tosatti} Santoro G, Scandolo S and Tosatti E  1999 {\it  \prb}{\bf 59} 1891\\
                  Anisimov V I, Bedin A E, Korotin M A, Santoro G, Scandolo S and
                  Tosatti E 2000 {\it \prb}{\bf 61} 1752
\bibitem{northrup} Northrup J and Neugebauer J 1995 {\it \prb}{\bf 52} R17001\\
                   Northrup J and Neugebauer J 1998  {\it \prb}{\bf 57} R4230
\bibitem{hellberg} Hellberg C S and Erwin S C  1999  {\it \prl}{\bf 83} 1003
\bibitem{wells} Wells B O, Shen Z X, Matsuura A, King D M, Kastner M A,
Greven M and Birgeneau R J 1995  {\it \prl}{\bf 74} 964
\bibitem{martinez} Martinez and Horsch P  1991  {\it \prb}{\bf 44} 317 \\
                   Liu Z and Manousakis  E 1992  {\it \prb}{\bf 45} 2425
\bibitem{kane} Kane C L, Lee P A and Read N  1989  {\it \prb}{\bf 39} 6880
\bibitem{azzouz} Azzouz M and Dombre Th 1996  {\it \prb}{\bf 53} 402 (1996)\\
                 Apel W, Everts H U  and K\"orner U  1998
                 {\it Eur. Phys. J.} B {\bf 5} 317 \\
                 Vojta M 1999 {\it \prb}{\bf 59} 6027
\bibitem{ramachandran99} Ramachandran V and Feenstra R M  1999  {\it \prl}{\bf 82} 1000
\bibitem{capriotti} Capone M, Capriotti L, Becca F and Caprara S  2001 {\it \prb}{\bf 63} 85104
\end{thebibliography}
\end{document}